\documentclass[sigconf,screen]{acmart} 

\usepackage{amsmath,amsfonts}

\usepackage{algorithmic}
\usepackage{booktabs}
\usepackage{array}
\usepackage{multirow}
\usepackage[utf8]{inputenc}
\usepackage{graphicx}
\usepackage{listings}
\usepackage{xcolor}
\usepackage{subcaption}
\usepackage{pifont}

\setcopyright{none}
\usepackage{xspace}
\newcommand*{\GH}{GitHub\@\xspace}
\newcommand*{\ie}{i.e.,\@\xspace}
\newcommand*{\eg}{e.g.,\@\xspace}

\newcommand{\rqone}{\textbf{RQ$_1$}: \emph{How well do LLMs support code obfuscation in terms of standard metrics?}}

\newcommand{\rqtwo}{\textbf{RQ$_2$}: \emph{How well can the semantic elasticity metric capture the ``obfuscation by simplification'' phenomenon?}}

\acmConference[EASE 2025]{The 29th International Conference on Evaluation and Assessment in Software Engineering}{17–20 June, 2025}{Istanbul, Turkey}

\begin{document}

\title{Simplicity by Obfuscation: Evaluating LLM-Driven Code Transformation with Semantic Elasticity}


\author{Lorenzo De Tomasi}
\email{lorenzo.detomasi@graduate.univaq.it}
\affiliation{%
\institution{University of L'Aquila}
\city{67100 L'Aquila}
\country{Italy}
}

\author{Claudio Di Sipio}
\email{claudio.disipio@univaq.it}
\orcid{0000-0001-9872-9542}
\affiliation{%
	\institution{University of L'Aquila}
	\city{67100 L'Aquila}
	\country{Italy}
}

\author{Antinisca Di Marco}
\email{antinisca.dimarco@univaq.it}
\orcid{0000-0001-7214-9945}
\affiliation{%
	\institution{University of L'Aquila}
	\city{67100 L'Aquila}
	\country{Italy}
}

\author{Phuong T. Nguyen}
\email{phuong.nguyen@univaq.it}
\orcid{0000-0002-3666-4162}
\affiliation{%
	\institution{University of L'Aquila}
	\city{67100 L'Aquila}
	\country{Italy}
}

\sloppy

\begin{abstract}

Code obfuscation is the conversion of original source code into a functionally equivalent but less readable form, aiming to prevent 
reverse engineering and intellectual property theft. 
This is a challenging task since it is crucial to
maintain functional correctness of the code while substantially disguising the input code. 
The recent development of large language models (LLMs) paves the way for 
practical applications 
in different domains, including software engineering. This work performs an empirical study on the ability of LLMs to obfuscate Python source code and introduces a metric (\ie semantic elasticity) to measure the quality degree of obfuscated code. 
We experimented with 
3 leading LLMs, \ie Claude-3.5-Sonnet, Gemini-1.5, GPT-4-Turbo across 30 Python functions from diverse computational domains.  
Our findings reveal GPT-4-Turbo's remarkable effectiveness with few-shot prompting (81\% pass rate versus 29\% standard prompting), significantly outperforming both Gemini-1.5 (39\%) and Claude-3.5-Sonnet (30\%). Notably, we discovered a counter-intuitive ``obfuscation by simplification'' phenomenon where models consistently reduce rather than increase cyclomatic complexity.
This study provides a methodological framework for evaluating AI-driven obfuscation while highlighting promising directions for leveraging LLMs in software security. 
\end{abstract}

\maketitle

\section{Introduction}
Protecting proprietary code has become a critical challenge for organizations. Code obfuscation is the transformation of code into a functionally equivalent but less readable form~\cite{collberg, low2017review}, serving as a 
defensive measure against reverse engineering and theft of intellectual property. 

Traditional code obfuscation employs several established approaches, i.e., layout transformations that alter formatting~\cite{laszlo2013}, control flow transformations that restructure logical paths~\cite{udupa2005deobfuscation}, data transformations that modify storage patterns~\cite{mohsen2017performance}, and identifier renaming that replaces meaningful names with arbitrary strings~\cite{low2017review}. Even though those techniques can be effective in practice, they often require manual adaptation to specific code patterns and can inadvertently introduce functional bugs~\cite{collberg}. Moreover, traditional metrics often focus solely on increasing complexity, which fails to capture the effectiveness of certain transformation techniques.


Large language models (LLMs) have paved the way for various software engineering applications \cite{10.1145/3695988}. These advanced tools opened new possibilities for automated code transformation~\cite{chen2023ai, golovko2024}, and demonstrate significant capabilities in understanding code semantics, logical structure, and programming idioms—qualities that make them potentially valuable tools for code obfuscation. However, the application of LLMs to security-oriented code transformation tasks remains relatively unexplored, with important questions about their effectiveness and optimal usage patterns.

In this paper, we present an empirical study 
of how modern LLMs can be applied to code obfuscation while preserving functional integrity. We experiment with 3 
models—Claude-3.5-Sonnet~\cite{anthropic2024claude}, Gemini-1.5~\cite{google2024gemini}, and GPT-4-Turbo~\cite{openai2024gpt4}—on a diverse dataset of 30 Python functions spanning five domains and 2 standard prompt engineering techniques 
\ie zero and few-shots.  
In addition, we propose a novel metric, called Semantic Elasticity $(SE)$, to overcome the limitations of the existing metrics. In particular, $SE$ quantifies a model's ability to radically transform code structure while maintaining functionality~\cite{behera2015complexity}.


We came across various interesting findings. First, we discovered the phenomenon "obfuscation by simplification", \ie LLMs tend to reduce code complexity rather than increase it when performing obfuscation, which contradicts traditional approaches. Second, we found out that 
GPT-4-Turbo obtains substantial improvement with few-shot prompting (80.93\% pass rate compared to 28.55\% with standard prompting); Gemini-1.5 demonstrates good processing efficiency; while Claude-3.5-Sonnet performs well in identifier transformations that effectively mask code intent. Last, we introduced ``Semantic Elasticity'' as a new evaluation metric that allows for a more flexible comparison for the code obfuscation task.  

To our knowledge, this is the first study ever to have investigated the application of LLMs in code obfuscation. In this respect, the main contributions of this paper are as follows:
 \textit{(i)} A quantitative analysis across three LLMs and six established metrics, conducted on a dataset spanning diverse application domains, revealing model-specific characteristics and effective usage patterns; \textit{(ii)} Identification of an "obfuscation by simplification" approach that differs from traditional approaches by decreasing rather than increasing code complexity while maintaining protection effectiveness, and \textit{(iii)} an initial exploration of a conceptual framework with the corresponding source code \cite{replication-package} for fostering further research. 
\label{sec:introduction}

\section{Motivation and Background}

\subsection{Problem statement}
Code obfuscation transforms readable code into functionally equivalent but difficult-to-understand versions. For instance, a factorial function and its obfuscated counterpart are represented in Listing \ref{lst:original} and Listing \ref{lst:obfuscated}.

\begin{lstlisting}[caption=Original factorial function,label=lst:original, language=Python]
def factorial(n):
    if n <= 1:
        return 1
    return n * factorial(n-1)
\end{lstlisting}

\begin{lstlisting}[caption=Obfuscated version,label=lst:obfuscated,language=Python]
def _I1l0O(n):
    def _calculate(x):
        return 1 if x <= 1 else x * _calculate(x-1)
    _unused = "x" * (n % 2)
    return _calculate(n)
\end{lstlisting}

The obfuscated version maintains identical functionality but incorporates confusing naming, nested helpers, and unnecessary variables—demonstrating the potential for LLMs to apply sophisticated transformations. More rigorously, we can formalize the problem 
following similar decomposition approaches in program analysis literature \cite{ming2015program}. For any model $M_i$, we can decompose the obfuscation operator as:
\begin{equation}
O_{M_i} = \alpha_i O_{struct} + \beta_i O_{name} + \gamma_i O_{logic} + \delta_i O_{rand}
\end{equation}
where $O_{struct}$ represents structural transformations, $O_{name}$ represents naming transformations, $O_{logic}$ represents logical transformations, and $O_{rand}$ represents randomization elements.

\subsection{Open challenges}

Despite extensive research, several challenges remain unresolved in code obfuscation. 
The functionality-obfuscation tradeoff represents a fundamental tension between making code difficult to understand while preserving correct functionality \cite{xu2018method}. Aggressive obfuscation often leads to runtime errors, creating a practical ceiling on transformation degree.

Traditional obfuscation tools typically apply fixed transformation patterns that adapt poorly to different code structures, reducing effectiveness against sophisticated reverse engineering attempts using machine learning techniques \cite{ceccato2017understanding}.

Quantification challenges persist without universally accepted metrics for evaluating how difficult obfuscated code is to understand \cite{anckaert2007measuring, xiao2022}. Traditional metrics like cyclomatic complexity~\cite{mccabe} may not fully capture comprehension difficulty.

Machine learning-based deobfuscation tools require increasingly sophisticated obfuscation techniques that can resist automated analysis \cite{haq2019survey, paloalto2024}, creating an ongoing arms race. 
Performance overhead remains a significant concern, as heavily obfuscated code often suffers from substantial performance penalties \cite{mohsen2017performance}, creating tension between security needs and operational requirements.
\label{sec:background}

\section{Evaluation methods and materials}
\begin{figure}
    \centering
    \includegraphics[width=0.9\linewidth]{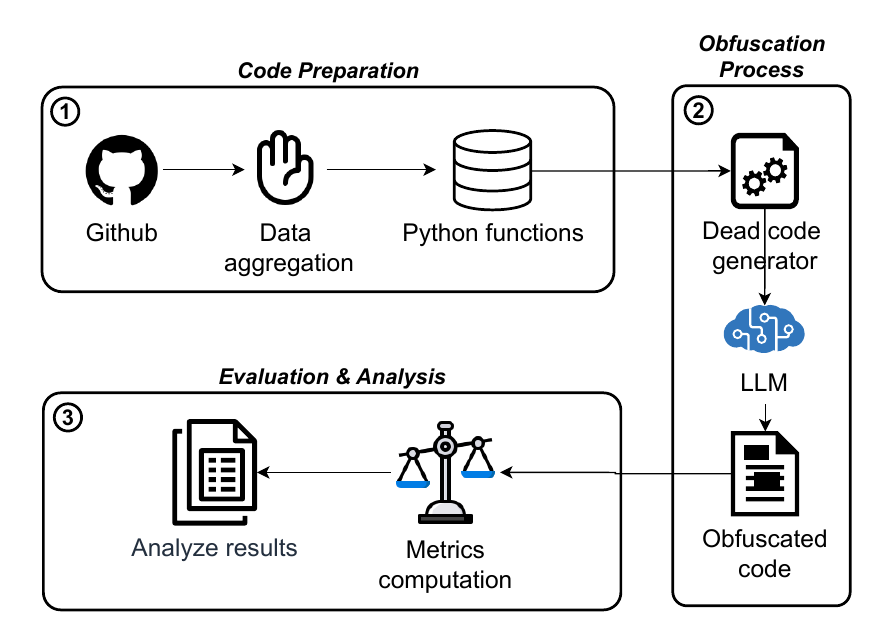}
    \caption{Proposed methodology.}
    \label{fig:approach}
\end{figure}

Figure \ref{fig:approach} depicts the proposed methodology. In particular, we first curate a dataset of Python functions in the \textit{code preparation} phase. Afterward, we run the \textit{obfuscation process} with different LLMs to obtain the obfuscated version of each snippet. We eventually analyze the results and compute state-of-the-art metrics discussed in Section \ref{sec:metrics} in the \textit{evaluation and analysis} phase.

Our investigation is centered on two main research questions:

\noindent \ding{226} \rqone~To answer this question, we curated a dataset composed of 30 Python functions belonging to five different categories defined in Section \ref{sec:dataset} to compare the selected models in terms of widely adopted metrics. In addition, we examined how the different categories can impact the overall obfuscation process. 

\noindent \ding{226} \rqtwo~We 
further analyze the "obfuscation by simplification” phenomenon measured in the RQ$_1$ by relying on the semantic elasticity metric defined in this paper.

\subsection{Dataset}
\label{sec:dataset}

We carefully selected 30 Python functions that are publicly available on \GH. Table~\ref{tab:categories_functions} summarizes the collected dataset. The elicited functions belong to the following five different categories.

\noindent \ding{226} \textit{Mathematical functions} represent fundamental computational algorithms spanning from basic to recursive calculations, \eg  factorial or fibonacci.
These functions exhibit diverse recursion patterns and computational complexities, providing baseline cases for evaluating the models' ability to preserve mathematical precision and optimization.

\noindent \ding{226} \textit{Sorting and searching algorithms} encompass both elementary implementations and more sophisticated approaches. These algorithms feature complex control structures with varying iteration and recursion patterns, making them ideal for testing transformations that must preserve execution order and comparative logic.

\noindent \ding{226} \textit{String manipulation techniques} include basic operations like character reversal and palindrome checking, as well as more complex algorithms such as Levenshtein distance and common substring identification. These functions manipulate immutable data types with diverse control flows and represent common use cases in text processing applications.

\noindent \ding{226} \textit{Data structure operations} focus on collection transformations and composite data type manipulation. These functions demonstrate how LLMs handle operations on nested and complex data structures, which are critical for evaluating obfuscation in data organization. 

\noindent \ding{226} \textit{Recursive algorithms} comprise classic problems like Tower of Hanoi and dynamic programming implementations such as the knapsack problem. These algorithms present deeply nested control structures and complex 
patterns, challenging the models with transformations that must preserve self-referential relationships.

\begin{table}[htbp]
\centering
\scriptsize
\footnotesize
\setlength{\tabcolsep}{2pt}
\renewcommand{\arraystretch}{1.1}
\caption{Overview of Categories and Functions.}
\begin{tabular}{|p{1.5cm}|p{4.8cm}|p{1.5cm}|}
\hline
\textbf{Category} & \textbf{Functions} & \textbf{Source} \\
\hline
Mathematical & factorial, fibonacci, is\_prime, gcd, lcm, power, sqrt\_newton & TheAlg.~\cite{thealgorithms2023} \\
\hline
Sorting \& Searching & bubble\_sort, insertion\_sort, quick\_sort, merge\_sort, binary\_search, linear\_search & TheAlg.~\cite{thealgorithms2023} \\
\hline
String Manipulation & str\_reverse, is\_palindrome, word\_count, levenshtein\_distance, longest\_common\_substring, count\_vowels & RuneSt.~\cite{runestone} \\
\hline
Data Structures & flatten\_list, dict\_merge, remove\_duplicates, rotate\_array, list\_permutations & PyPatt.~\cite{faif2023} \\
\hline
Recursive & tower\_of\_hanoi, binary\_tree\_depth, flood\_fill, knapsack, edit\_distance, coin\_change & RuneSt.~\cite{runestone} \\
\hline
\end{tabular}
\label{tab:categories_functions}
\end{table}

 
We selected functions that represent common programming paradigms, varying complexity levels, computational patterns, diverse control flow structures, and different parameter types, ensuring adequate coverage of real-world programming functions.

We use three different code repositories to avoid any bias in the selection:
\ie \textit{TheAlgorithms/Python~\cite{thealgorithms2023}}, \textit{python-patterns~\cite{faif2023}}, and 
\textit{Problem-Solving with Algorithms and Data Structures~\cite{runestone}} 



For each function, we generated comprehensive test cases specifically tailored to verify functional equivalence between original and obfuscated implementations. The test suite included edge cases, typical usage patterns, and stress tests to ensure a robust 
validation.

\subsection{Metrics}
\label{sec:metrics}

To rigorously evaluate the obfuscation performance, we employed a comprehensive set of complementary metrics as follows~\cite{collberg2002watermarking,ceccato2014family}. 

The \textbf{Pass Rate} $(P)$ measures the percentage of test cases where obfuscated code maintains functional equivalence with the original—arguably the most critical aspect of successful obfuscation as shown in Equation \ref{eq:pass}. In addition, it has been used in prior work to assess LLM capabilities~\cite{10.1145/3690635}. For each function, we executed both original and obfuscated versions against identical test cases, comparing outputs for exact equality.
\begin{equation}\label{eq:pass}
P = \frac{1}{n}\sum_{i=1}^{n} \mathbb{I}(f_{\text{orig}}(t_i) = f_{\text{obfs}}(t_i))
\end{equation}
where $n$ is the total number of test cases, $t_i$ is the $i$-th test case, $f_{\text{orig}}(t_i)$ is the output of the original function for test case $t_i$, $f_{\text{obfs}}(t_i)$ is the output of the obfuscated function for test case $t_i$, and $\mathbb{I}(x)$ is the indicator function that equals 1 when $x$ is true and 0 otherwise.

\textbf{Code expansion ratio} $(E)$ captures the ratio of lines in the obfuscated function compared to the original~\cite{748622}, calculated as described in Equation \ref{eq:codeExp}:
\begin{equation}\label{eq:codeExp}
E = \frac{\text{lines in obfuscated code}}{\text{lines in original code}}
\end{equation}
\textbf{Cyclomatic complexity change} $(\Delta CC)$ measures the difference in McCabe's cyclomatic complexity~\cite{mccabe,gill2011cyclomatic}, calculated by counting decision points in the abstract syntax tree (AST) (see Equation \ref{eq:cc}. We implemented this using Python's \texttt{ast} module with a custom visitor pattern to count branches. Our implementation counted \texttt{if}, \texttt{while}, \texttt{for}, and \texttt{lambda} expressions as complexity points as follows.
\begin{equation}\label{eq:cc}
\Delta CC = CC_{\text{obfuscated}} - CC_{\text{original}}
\end{equation}
\textbf{Identifier entropy change} $(\Delta H)$ quantifies the change in Shannon entropy~\cite{shannon,gopan2007string} of variable names, with higher entropy indicating more random, less meaningful identifiers that are harder to interpret. We compute the entropy following Equation \ref{eq:entropy}:
\begin{equation}\label{eq:entropy}
H = -\sum_{i=1}^{n} p_i \log_2 p_i
\end{equation}
where $p_i$ is the frequency of the $i$-th identifier in the code, and:
\begin{equation}
\Delta H = H_{\text{obfuscated}} - H_{\text{original}}
\end{equation}
A positive value of $\Delta H$ indicates increased randomness and decreased comprehensibility of identifiers in the obfuscated code. Higher $\Delta H$ values suggest that variable and function names have become more uniformly distributed and less semantically meaningful, making it harder for an analyst to infer program behavior from identifier inspection alone.

\textbf{Execution time difference} $(\Delta T)$ measures runtime performance differences between original and obfuscated versions~\cite{udupa2005deobfuscation,ahmadvand2018taxonomy}, calculated as the average difference across all test cases as defined in Equation \ref{eq:execTime}:
\begin{equation}\label{eq:execTime}
\Delta T = \frac{1}{n}\sum_{i=1}^{n}(T_{\text{obfuscated},i} - T_{\text{original},i})
\end{equation}


\noindent \textbf{Semantic Elasticity} As discussed in Section \ref{sec:introduction}, the abovementioned traditional metrics focus only on the complexity of the obfuscated code. To overcome this limit, we proposed the Semantic Elasticity metric defined in Equation \ref{eq:se}: 



\begin{equation}\label{eq:se}
SE = \frac{|\Delta CC| \times P^2}{E}
\end{equation}

Roughly speaking, we follow similar approaches to combined software metrics \cite{olague2008comprehensive} to define $SE$ metric by integrating the absolute cyclomatic complexity change $(|\Delta CC|)$, the square of pass rate $(P^2)$ to emphasize functional correctness, and inversely relates to code expansion $(E)$ to reward efficiency~\cite{potdar2004comprehensive}.

Semantic Elasticity provides several advantages over traditional metrics. First, it captures effective transformations regardless of whether they increase or decrease complexity~\cite{anckaert2007proteus}. Second, it emphasizes the critical importance of maintaining functionality through the squared pass rate term. Third, it rewards transformations that achieve significant structural changes with minimal code expansion. Finally, it provides a single, holistic measure that balances multiple aspects of obfuscation quality~\cite{viticchie2016metrics}.


\label{sec:methods}

\section{Experimental Results}


\subsection{Answering \textbf{RQ$_1$}}

\begin{figure}[!t]
\centering\includegraphics[width=0.9\linewidth]
{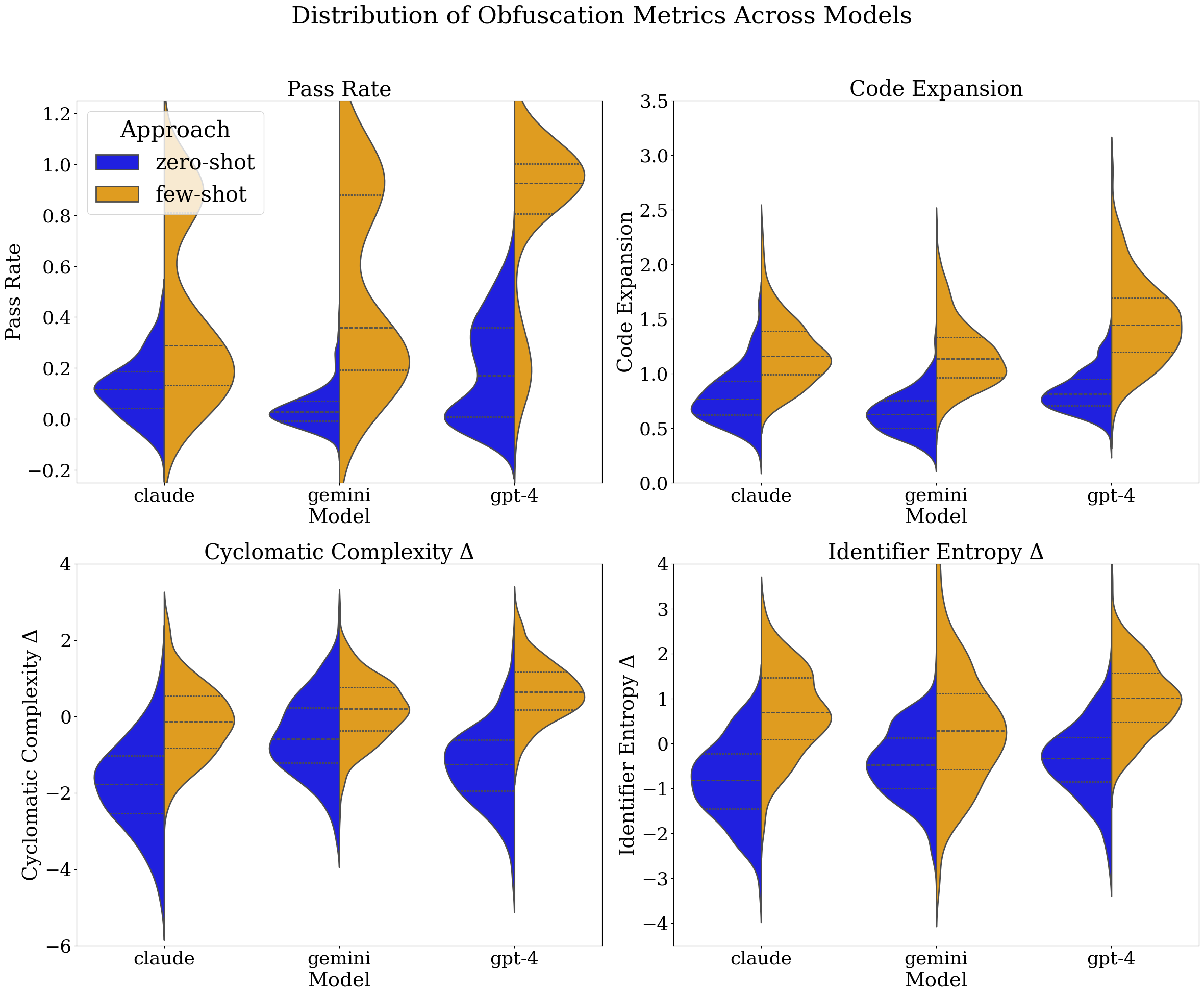}
\caption{Distribution of obfuscation metrics across models. 
}
\label{fig:violin_plots}
\end{figure}

Figure~\ref{fig:violin_plots} depicts the results obtained by the examined models in terms of the standard obfuscation metrics.  



It is evident that GPT-4-Turbo outperforms the other models when instructed with few-shot prompting, achieving an impressive 80.93\% pass rate, far exceeding both Gemini-1.5 (39.23\%) and Claude-3.5-Sonnet (29.97\%). Under standard prompting, GPT-4-Turbo maintained its lead with a 28.55\% pass rate, while both Claude (9.00\%) and Gemini (2.01\%) struggled significantly. As shown in Figure~\ref{fig:violin_plots} (top-left), the pass rate distributions reveal a striking bimodal pattern for GPT-4 in few-shot mode, indicating a clear separation between successful cases (clustered near 1.0) and failures (near 0.0), rather than the gradual distribution one might expect.

For code expansion, all models showed moderate growth under few-shot prompting, with GPT-4-Turbo showing the highest expansion ratio (1.004), followed by Gemini-1.5 (0.811) and Claude-3.5-Sonnet (0.803). Standard prompting produced less expansion across all models. 
We also report that the actual distributions contain significant outliers, with some obfuscated solutions expanding to over 4 times the original size, particularly for Claude and GPT-4.



Unexpectedly, all models consistently reduced rather than increased cyclomatic complexity, contrary to traditional obfuscation techniques that typically aim to inflate complexity \cite{udupa2005deobfuscation}. Claude-3.5-Sonnet showed the most pronounced reduction in standard prompting (-1.27), followed by GPT-4-Turbo (-1.03) and Gemini-1.5 (-0.47). Notably, the violin plots reveal extreme cases where complexity was reduced by up to 6 points, especially with Claude and Gemini.

For identifier entropy changes, GPT-4-Turbo demonstrated the most effective approach in few-shot prompting (+0.52), while Claude-3.5-Sonnet achieved positive changes (+0.23). Interestingly, all models showed negative entropy changes under standard zero-shot prompting, suggesting they struggled to implement effective identifier obfuscation without examples.




Statistical analysis confirms significant differences between standard and few-shot approaches. T-tests show statistically significant differences in pass rates ($t=-13.12$, $p<0.0001$), complexity change ($t=-2.50$, $p=0.0125$), and entropy change ($t=-2.80$, $p=0.0052$). This confirms that prompt engineering substantially impacts obfuscation effectiveness.

Function category analysis reveals that String Manipulation functions achieved the highest pass rate (55.82\%), followed by Data Structure (54.67\%), while Recursive algorithms showed the lowest (42.16\%). This pattern suggests that code obfuscation using few-shots PET is more effective for certain functions involving algorithmic structures, while recursive logic patterns are more challenging.

Few-shot prompting dramatically shifts performance patterns across all models, as shown in Figure~\ref{fig:violin_plots}. The overall improvement in pass rates from standard to few-shot prompting was substantial, with the combined average increasing from 13.19\% to 50.04\%.

Our statistical analysis confirms the significance of these differences. The $t$-test comparing standard and few-shot pass rates yielded a $t$-statistic of -13.12 and a $p$-value less than 0.0001, providing strong statistical evidence that the prompt engineering approach substantially impacts obfuscation performance. Similar significant differences were found for complexity change ($p=0.0125$) and entropy change ($p=0.0052$).

Furthermore, we analyze the impact of the few-shot PET on the various function categories. As shown in Table~\ref{tab:categories}, all function categories showed dramatic improvements with few-shot prompting but with varying magnitudes. The most substantial gains were seen in String Manipulation functions (44.58\%) and Data Structure functions (43.03\%).

The varying response to few-shot prompting across function categories suggests that certain algorithmic patterns benefit more from example-guided transformation than others. String Manipulation functions, which often involve straightforward transformations with clear input-output relationships, showed the highest few-shot success rates. In contrast, Recursive algorithms, with their more complex control flow and state management, proved more challenging despite the availability of examples.

\begin{table}[h]
\caption{Pass-rated results for each function category.}
\label{tab:categories}
\footnotesize
\begin{tabular}{l c c}
\toprule
\textbf{Category} & \textbf{Standard} & \textbf{Few-shot} \\
\midrule
Data Structure & 11.64\% & 54.67\% \\
Mathematical & 19.25\% & 52.52\% \\
Recursive & 8.11\% & 42.16\% \\
Sorting \& Searching & 14.56\% & 48.44\% \\
String Manipulation & 11.24\% & 55.82\% \\
\bottomrule
\end{tabular}
\end{table}

\noindent\fbox{\begin{minipage}{0.98\columnwidth}
		\paragraph{\textbf{Answer to RQ$_1$:}} 
        Claude-3.5-Sonnet prioritizes naming transformations ($\beta$) and structural reimagining. Meanwhile, Gemini-1.5 maintains a balanced approach across transformation types, and GPT-4-Turbo emphasizes logical transformations ($\gamma$). 
        Code obfuscation strongly depends on the type of function considered.  
\end{minipage}}

\subsection{Answering \textbf{RQ$_2$}}

\begin{figure}
    \centering
    \includegraphics[width=0.8\linewidth]{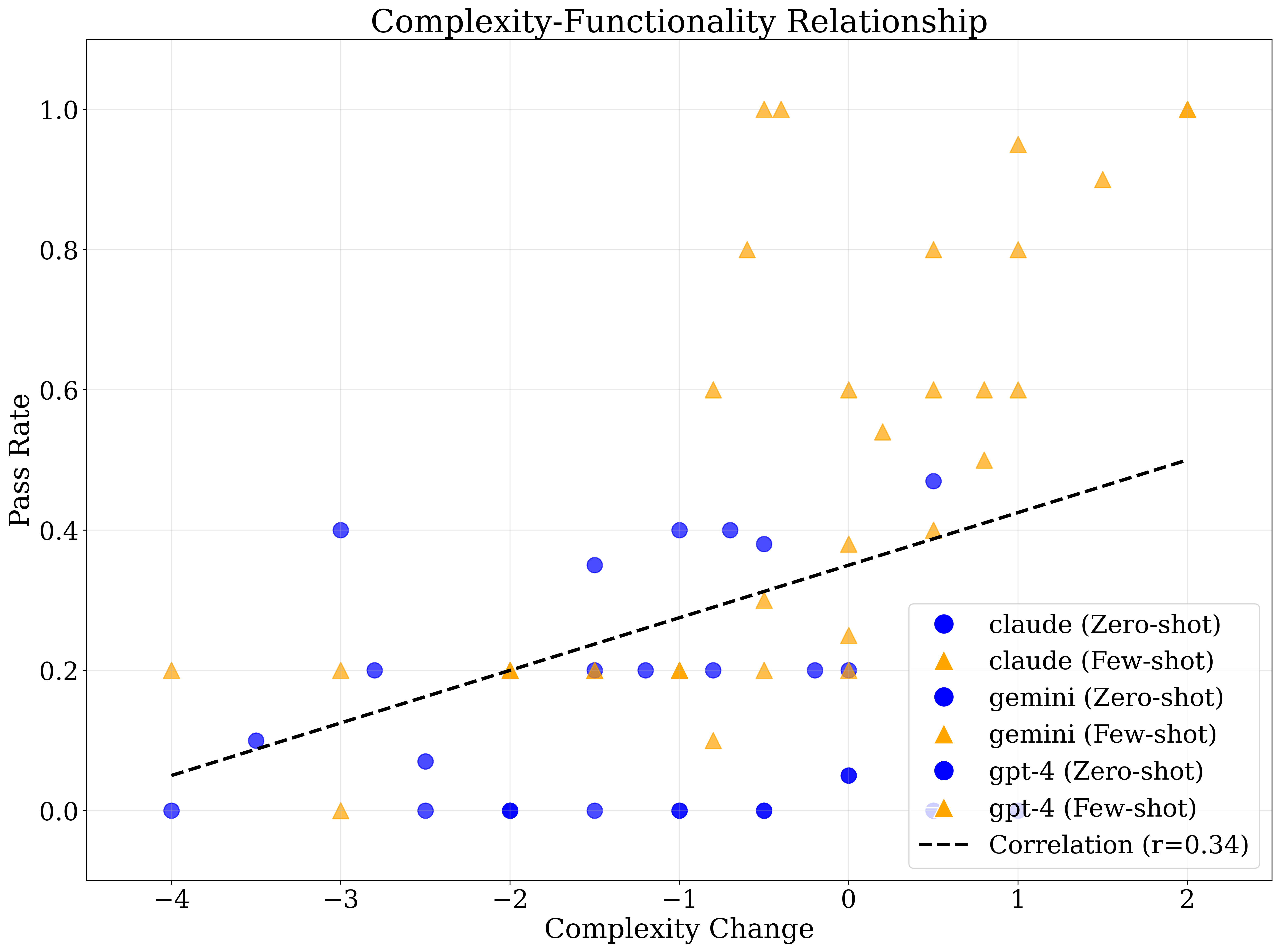}
    \caption{Correlation analysis.}
    \label{fig:corr}
\end{figure}
\begin{figure}
    \centering
    \includegraphics[width=0.8\linewidth]{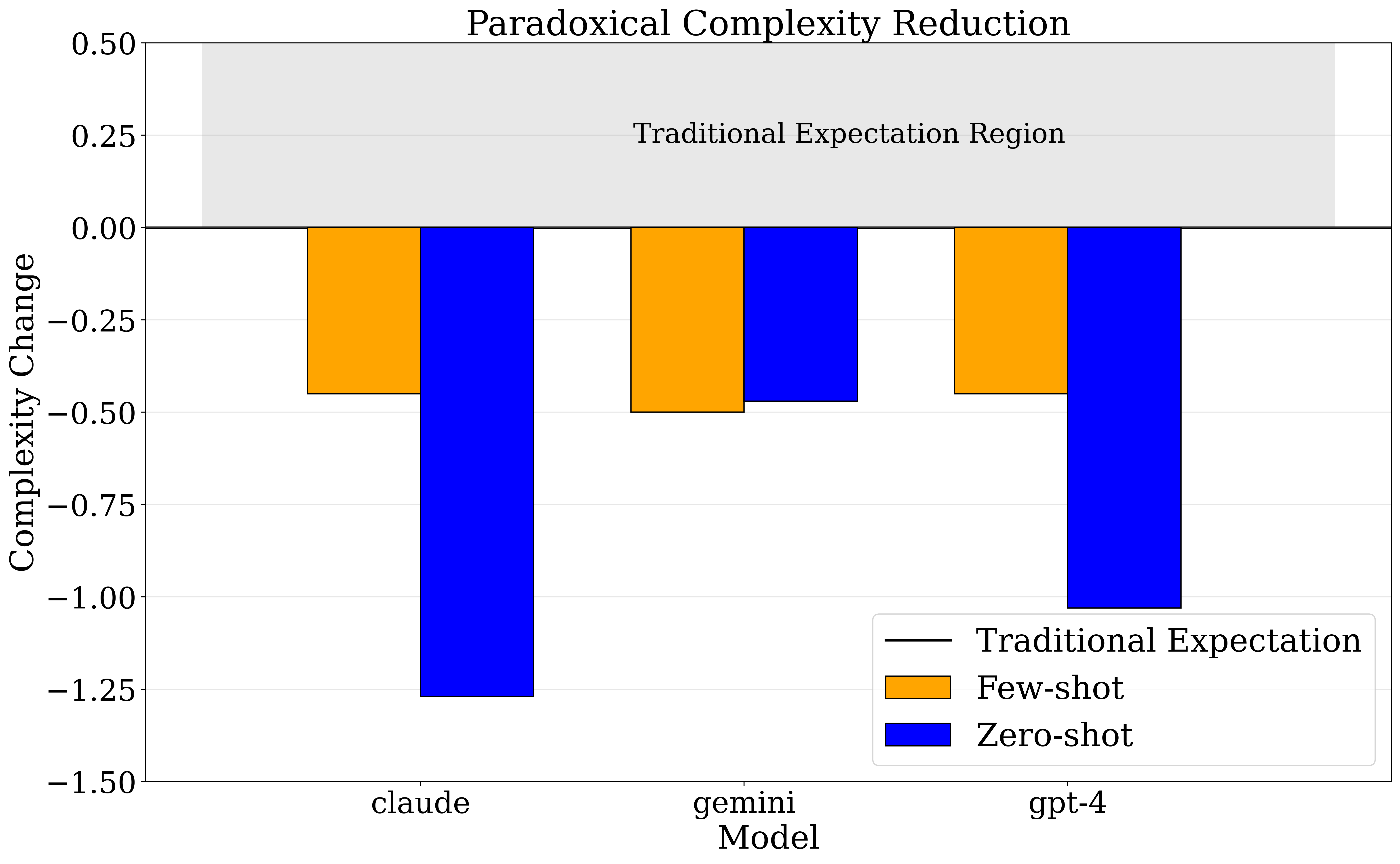}
    \caption{Complexity reduction analysis.}
    \label{fig:complex}
\end{figure}

To study if the examined LLMs can simplify the obfuscated code, we further analyze the obtained results by computing correlation. As shown in Figure~ \ref{fig:complex}, the majority of obfuscation attempts resulted in reduced cyclomatic complexity in all models and approaches. Overall, the models consistently reduces complexity, and the standard approaches showed more dramatic reductions (-0.91 average) than the few shot approaches (-0.66 average). In addition, we conducted a correlation analysis shown in Figure \ref{fig:corr}. Our analysis reveals a positive correlation between complexity change and pass rate, suggesting that \emph{excessive simplification may compromise functionality.}

In addition, we compute the Semantic Elasticity metric for each model instructed with the two prompt techniques as summarized in Table \ref{tab:semantic_elasticity}. Overall, zero-shot PET outperforms the few-shot for all the examined models except for Claude. GPT-4-Turbo achieves the best result compared to the others, with a 0.730 value on average. Therefore, we observed the "simplicity by obfuscation" phenomenon that seems to not compromise the coding efficiency like traditional techniques \cite{mohsen2017performance}.

This could indicate its effectiveness at balancing structural transformation with functionality preservation, although further analysis is needed to confirm our intuition. 

\begin{table}[htbp]
\centering
\footnotesize
\caption{Semantic Elasticity results.}
\begin{tabular}{|l|l|c|c|c|c|}
\hline
\textbf{Model} & \textbf{PET} & \textbf{Mean} & \textbf{Std Dev} & \textbf{Min} & \textbf{Max} \\ \hline
\multirow{2}{*}{GPT-4} & Zero-shot & \textbf{0.730} & 1.876 & 0.000 & \textbf{15.429} \\ \cline{2-6}
                       & Few-shot & 0.162 & 0.544 & 0.000 & 3.000 \\ \hline
\multirow{2}{*}{Claude} & Zero-shot & 0.061 & 0.242 & 0.000 & 1.512 \\ \cline{2-6}
                        & Few-shot & \textbf{0.182} & 0.381 & 0.000 & 2.000 \\ \hline
\multirow{2}{*}{Gemini} & Zero-shot & \textbf{0.178} & 0.801 & 0.000 & 4.600 \\ \cline{2-6}
                        & Few-shot & 0.038 & 0.149 & 0.000 & 0.727 \\ \hline
\end{tabular}
\label{tab:semantic_elasticity}
\end{table}

\noindent\fbox{\begin{minipage}{0.98\columnwidth}
		\paragraph{\textbf{Answer to RQ$_2$:}} 
        Our experiment reveals that LLMs are capable of simplifying the obfuscated code in terms of complexity, reducing rather than increasing it. In addition, the semantic elasticity seems to be maximized with standard zero-shot prompting, even though in-depth analyses are required to confirm this outcome. 
\end{minipage}}

\section{Threats to validity}

\textbf{Internal validity} concerns our experimental setup. Although we included 30 diverse functions from established repositories, they may not represent all programming patterns. Even though the selected models represent the current state-of-the-art, we left out open-source LLM from our analysis since we focus on proprietary ones for security issues. Nevertheless, our methodology can be applied to every LLM architecture as we did not use any peculiar feature of the considered LLMs. Our metrics, while comprehensive, cannot capture all aspects of obfuscation quality. The proposed Semantic Elasticity metric has not been validated across different contexts and coding paradigms and should be considered an initial exploration rather than a definitive measurement.

\textbf{External validity} relates to generalizability. Our focus on Python code may limit applicability to other languages with different paradigms or syntax \cite{balakrishnan2018}. Real-world applications often involve complex interactions between multiple components, while our evaluation focused on isolated functions \cite{herraiz2011beyond}. We evaluated obfuscation effectiveness primarily through code structure metrics rather than resistance to actual deobfuscation attempts, which could provide a more complete assessment of protection effectiveness.
\label{sec:results}


\section{Related Work}

Collberg and Nagra~\cite{collberg} provided a comprehensive taxonomy of obfuscation techniques, categorizing them into layout, control flow, and data transformations. 
After defining a set of metrics to evaluate obfuscation, they evaluated Java bytecode transformations with quantitative measurements of complexity increases and performance penalties across various algorithms, stating that effective obfuscation typically increases computational overhead by 10-30\%.

Laszlo et al.~\cite{laszlo2013} proposed methods for measuring obscurity using control flow complexity and structural entropy metrics. Their evaluation methodology involved applying various transformations to C++ programs and quantifying the increase in control-flow graph (CFG) complexity through both static analysis and user studies with 25 programmers attempting to understand the transformed code. 

Ceccato et al.~\cite{ceccato2017understanding} conducted a comprehensive empirical study with 30 professional hackers tasked with attacking protected and unprotected software, using mixed-methods analysis combining quantitative metrics with qualitative analysis of hacker behavior through think-aloud protocols. Their findings revealed that even professional tools could be circumvented with an average 2.8x increase in attack time compared to unprotected code.



Chen et al.~\cite{chen2018automated} explored automated vulnerability detection using deep representation learning. They evaluated their approach on a dataset of 4,122 software functions from open-source projects, using a combination of static analysis and deep neural networks. Their method achieved 89\% accuracy in detecting buffer overflow vulnerabilities, demonstrating the potential of deep learning for understanding complex code semantics.

Our work offers the first evaluation of state-of-the-art LLMs for code obfuscation across 30 diverse functions and five different categories. In addition, we introduced "Semantic Elasticity" as a novel metric integrating functionality preservation with transformation effectiveness. 

\label{sec:related}

\section{Conclusion}

Motivated by the current limitations of existing code obfuscation techniques and evaluation metrics, this paper proposes an initial study that focuses on using Large Language Models (LLMs) for code obfuscation tasks for Python. By considering a curated dataset of different functions, we compared three prominent LLMs, \ie GPT-4 Turbo, Gemini, and Claude Sonnet, instructed with zero and few-shots prompting techniques. In addition, we propose a novel metric, called Semantic Elasticity, to overcome the limitations of the existing metrics.

Our analysis reveals that GPT-4-Turbo demonstrated excellent performance with few-shot prompting, achieving substantially higher pass rates compared to standard prompting approaches. We discovered the intriguing ``obfuscation by simplification'' phenomenon across all models, where effective code protection was achieved through complexity reduction rather than increase. Additionally, we observed that different algorithm type leads to different results in terms of the examined metrics, with string manipulation functions proving most amenable to successful transformation while recursive algorithms presented greater challenges.

For future work, we plan to devise enhanced prompt engineering techniques to further optimize the obfuscated code and hybrid approaches combining multiple models' complementary strengths may create more robust protection systems. 
We will extend our analysis to additional programming languages leveraging notable benchmarks, \eg Rosetta code\cite{rosettacode} or LeetCode \cite{leetcode}. 
Finally, we will further investigate the semantic elasticity concept with additional experiments.


\label{sec:conclusion}

\begin{acks} 
	
	This paper has been partially supported by the MOSAICO project (Management, Orchestration and Supervision of AI-agent COmmunities for reliable AI in software engineering) that has received funding from the European Union under the Horizon Research and Innovation Action (Grant Agreement No. 101189664). The work has been partially supported by the EMELIOT national research project, which has been funded by the MUR under the PRIN 2020 program (Contract 2020W3A5FY). It has been also partially supported by the European Union--NextGenerationEU through the Italian Ministry of University and Research, Projects PRIN 2022 PNRR \emph{``FRINGE: context-aware FaiRness engineerING in complex software systEms''} grant n. P2022553SL. We acknowledge the Italian ``PRIN 2022'' project TRex-SE: \emph{``Trustworthy Recommenders for Software Engineers,''} grant n. 2022LKJWHC. 
\end{acks}

\bibliographystyle{ACM-Reference-Format}
\bibliography{references}

\end{document}